\documentclass[notitlepage,letterpaper,11pt]{article} % para articulo en castellano
\usepackage[ansinew]{inputenc} % Acepta caracteres en castellano
\usepackage{amsmath}
\usepackage{amsfonts}
\usepackage{amssymb}
\usepackage[colorlinks=true,urlcolor=blue,linkcolor=blue]{hyperref} % navega por el doc
\usepackage{graphicx}
\usepackage{geometry}           % See geometry.pdf to learn the layout options.
\usepackage{epstopdf}
\usepackage{fancyhdr} % encabezados y pies de pg

\begin{document}
\title{ASTRONOMÍA AL AIRE:  \\
MEDIA CONVERGENCE \\ IN ASTRONOMY \& ASTROPHYSICS.
}
\author{
\textbf{L. A. Núñez and H. Rago} \\ 
\textit{Escuela de Física, Universidad Industrial de Santander, Bucaramanga-Colombia} and \\
\textit{Departamento de Física, Universidad de Los Andes, Mérida-Venezuela} \\
}%\date{} 
\maketitle                                          % Activate to display a given date or no date
\begin{abstract}
We describe the experience of running an Astrophysics outreach initiative involving traditional mass media like radio broadcast and new digital media like blog, microblogging and internet video channel. Some very successful preliminary results are also presented. This unique experience is helping to create new science informal education environments for Spanish speaking people.
\end{abstract}

\section{Media convergence and science outreach}
For science informal education environments, the convergence of radio, television, Internet and hand-held devices is pervasive,  making science information increasingly available to general public. Outreach science media is shaping qualitatively the people's relationship with science and is becoming an important new strategy for supporting science learning. Although there are strong evidences for the impact of educational television on science learning, substantially less evidence exists on the impact of other media -digital media, gaming, radio  \cite{FederEtal2009,StocklmayerRennieGilbert2010,SaccoFalkBell2014}.

The idea of digital media convergence, prophesied by several authors \cite{Negroponte1991,Fidler1997}, has been around for decades  (see  \cite{Mueller1999} and references therein) but it has been particularly boosted by the emerging informational economy \cite{Castells2001}.  Online television, web-radios, blogs and microblogging -with journalist, journals and news agencies as main actors- interplay among them, exchanging audience, reinforcing each other and building new scenarios on the comercial arena.

Recently, sprang by the availability of web 2.0 tools, individuals (mainly journalist) got the possibility to directly communicate their opinions to wider audience. Nowadays, scientist and scientific institutes are just starting to use blogs and microblogging to disseminate their research activities to the general public. Despite that there are very few studies about the impact and the sociology of digital science media ecosystem and emerging academic social networks \cite{VanNoorden2014,McclainNeeley2014,Brown2015}, there are clear examples of the use of these tools for science outreach mainly lead by big scientific consortiums \cite{GoldfarbKahleRao2014}.

Astronomía al Aire is an outreach initiative, part of science informal education project,  of the Relativity and Gravitation Research Group (GIRG for its Spanish acronym for Grupo de Investigación en Relatividad y Gravitaci\'on) and the School of Physics of Universidad Industrial de Santander, Bucaramanga-Colombia. Profiting from the natural fascination that Astronomy, the Universe and its occupants have on humanity imagination, it is aimed to reduce the gap between the specialised science and scientific conceptions of common citizens.  Astronomía al Aire is a unique experience that mixes traditional media like radio broadcast with blog and microblogging, discussing science topics related to Astronomy and Astrophysics and emphasising on significative values of science: its structure, its beauty and its essential atmosphere of creative freedom.

\section{Astronomía al Aire: \textit{the concept \& impact}}
In this section we describe the concept and some preliminary successful results of this multimedia outreach enterprise. 
We shall present how the interplay from traditional mass media like radio and emerging digital media like blog and microblogging is helping to create new science informal education environments in several countries of Latin-America.
 
\paragraph{Audio clips and podcasts}
Audio clips ($\sim$ 4 to 5 minutes) are the central part of Astronomía al Aire project and  currently 35 of them have been produced and disseminated weekly through our website\footnote{\url{http://halley.uis.edu.co/aire/}}. Based on previous experience we choose the clip author to be the storyteller narrator. The voice of a scientist that knows and understands the emphasis on content, is preferable and more credible than the neutral, trained voice of a professional broadcaster. 

The covered topics are diverse and, as a rule, we try to emphasise on the values of science: its structure, its beauty, its essential atmosphere of creative freedom. Clips focus permanently on the intimate relation  between theory and observations and are worked out about: scientists who have built theories and make discoveries, as well as relevant physical objects and/or event such as supernovae, stars, black holes, exoplanets, neutrinos, antimatter, the large structure of our universe. 

There are necessary overlapping of themes and subjects: clips on the evolution of stars shares concepts with others related to neutron stars, pulsars and/or black holes; micros on the expansion of the universe touch aspects with those of dark matter and dark energy.

Background music composition is unorthodox, avoiding every cliché relating science with any fuzzy electronic sound. Tango, salsa, Beatles, Stravinsky or Xenakis themes can help to create attractive atmosphere inviting the audience to get into a scientific message. 

\paragraph{Weekly radio program} 
A 30m weekly radio program complements audio clips presenting and discussing recent advances in Physics and Astrophysics. These longer programs also promote the activities of research groups from the School of Physics at the Universidad Industrial de Santander, including interviews to personalities related to our School.

\begin{figure}[!ht]
\begin{tabular}{cc}
	\includegraphics[width=0.48\textwidth]{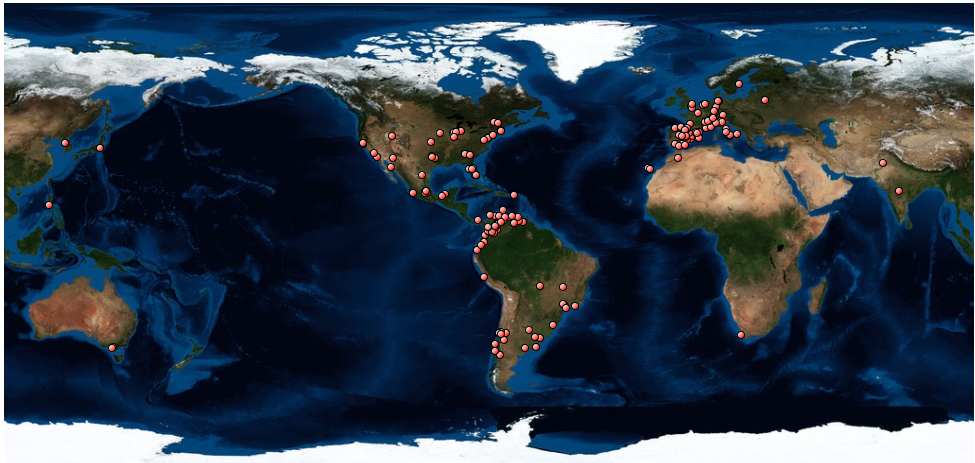} & 
	\includegraphics[width=0.48\textwidth]{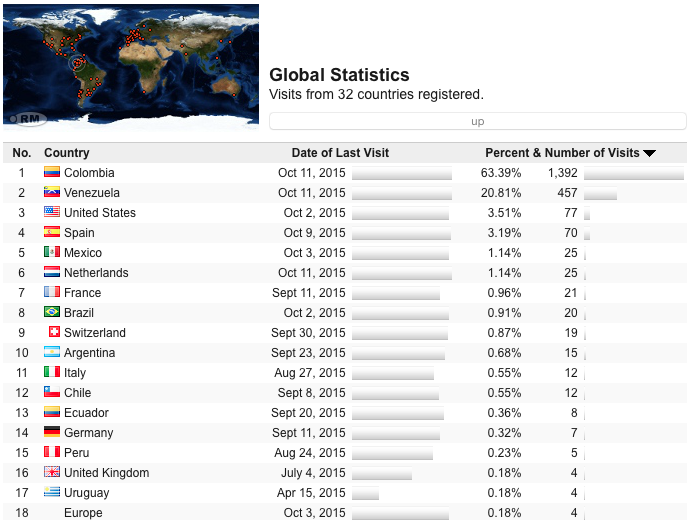} \\
	(a) & (b)
\end{tabular}
\caption{Our blog \url{http://halley.uis.edu.co/aire/} visited by more than 2500 times, from 32 countries in last seven months. (Snapshot taken with \url{https://www.revolvermaps.com} on Oct/11/2015)}
\label{BlogWorldVisit}
\end{figure}

\paragraph{Radio broadcasting}
Along this rationale, clips are regularly broadcast three times (7:30am, 12:30pm and 5:30pm) per day by 670Khertz AM UIS University radio in Bucaramanga and retransmitted in different schedules by several members of  RRULAC (a nine country Network of University Radio of Latin America and the Caribbean\footnote{\url{http://rrulac.org}}) in Mexico, Colombia and Venezuela.

\paragraph{The blogging}
Our blogsite\footnote{\url{http://halley.uis.edu.co/aire/}} designed to blend traditional and digital media has been visited from more than 2500 times, and 32 countries in last seven months and this can be appreciated from Figure \ref{BlogWorldVisit}). This website preserve and disseminate, text, clips and long radio productions and will include in the near future videos and wikibooks associated to selected of post and postcasts.

\begin{figure}
\begin{center}
\includegraphics[width=4in]{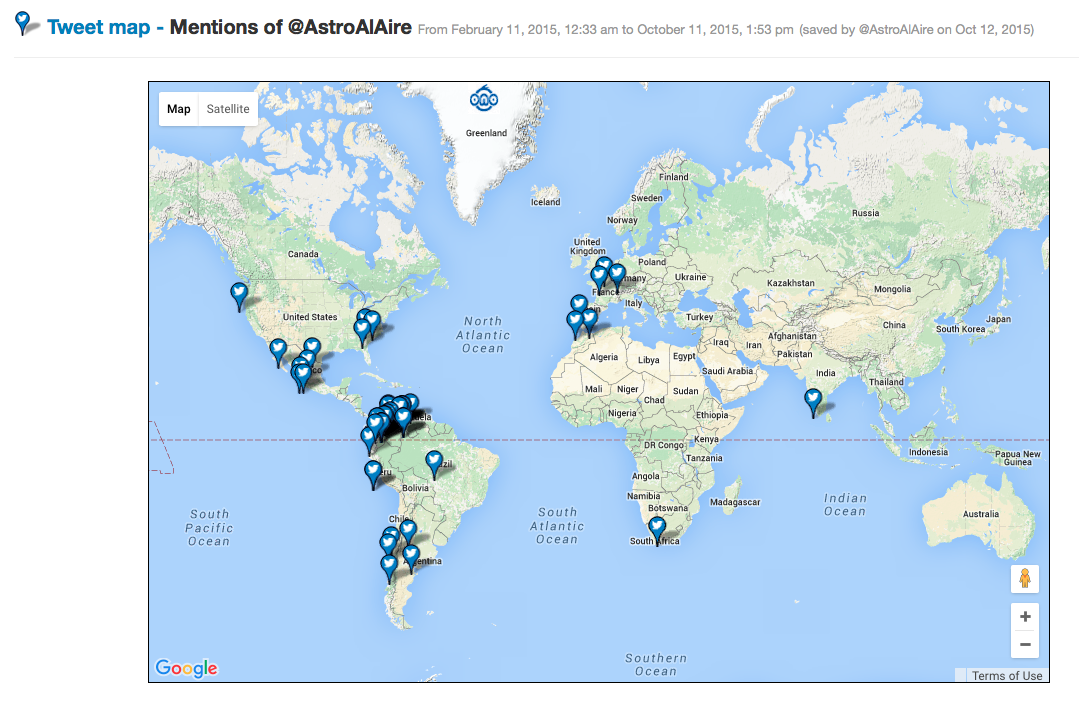}
\caption{@AstroAlAire in numbers from 11/02/2015 to 11/10/2015: 370 mentions, 289 favorited,  504 retweeted 917 times with 1.82 Average retweets per retweeted, with a potential impact on 2.128.330 readers (Snapshot taken with \url{http://www.twitonomy.com} on Oct/11/2015).  }
\label{AstroAlAireRT}
\end{center}
\end{figure}

\paragraph{The microblogging}
We create tweeter identity, @AstroAlAire, to foster astronomical and science contents to other audiences and, its natural dynamism rapidly expands the number of readers/listeners to our blog and radio clips. Presently, @AstroAlAire has fired around 2000 tweets in these seven months and it is followed by more than 400 other microbloggers. Several statistical tools allow us to trace the origin of: followers, mentions, retweets and favorites that summed up @AstroAlAire potential impact  to more than 2.000.000 followers. It is clear from Figure \ref{AstroAlAireRT} that we have found a significant impact of @AstroAlAire through the Andes in countries which conform the new Andean node of the Office of Astronomy for Development (OAD)\footnote{\url{http://www.astro4dev.org}}, i.e. Venezuela, Colombia, Ecuador, Perú, Bolivia and Chile. People from those countries retweet and favourite more frequently @AstroAlAire.

\section{Final remarks} 
In Latin America, as well as in many developing regions of the world, radio broadcasts have a tremendous penetration into general public audience, mainly for those who do not have access to the internet. In rural areas this penetration is more dramatic and important, becoming almost a public service to integrate geographically disperse and remote population. Within urban zones, the same scenario is reproduced: radio broadcast reach mostly laypeople with limited or non access to the internet very little informed about science. Thus, our intention it to captivate the interest of this population for Astronomy and Astrophysics message. 

Our weekly clips try to seduce and not to teach the audience, because in a short lapse of five minutes it is difficult to teach anything, but it is easy to capture the interest for a scientific inquire. To accomplish this task the metaphorical, poetically nice language is necessary, but it is more important to show that science has accurate codes and language that should be respected, even in communicating scientific concept to the general public.

Radio and press, in conjunction with blogs, microblogging and video channels, reinforcing each other,  become an effective mean to bridge the gap between scientific disciplines and other sectors of society.  Our next step will be twofold: first to incorporate wikibook than could be printed as pocket books with QR-code pointing to original posts and second, to create short videos illustrating selected audio clips. 

As far as we know Astronomía al Aire is a unique experience of media convergence delivering a scientific message to several segments of the general public, i.e. with and without regularly access to the Internet. Surely in Spanish there are no similar experiences and we did not find any trace of other alike initiative in French, Italian, Portuguese or English. Despite we are just going along the first year of this experience, and more time is needed to fully evaluate its impact,  the interest that our clips, blogs and microblogging have risen among the Spanish speaking university radios is encouraging.

It is of a paramount importance to foster these other sectors to understand the nature of science, research and how they are worked out in their daily lives. Presently, the convergence of digital and nondigital media has the potential of assisting the creation of a scientific culture and its important potential should be exploited.

\section*{Acknowledgments} 
We acknowledge the financial support of Vicerrector\'{\i}a de Investigaci\'on y Extensi\'on and Vicerrector\'{\i}a Administrativa from Universidad Industrial de Santander. We also thank Jorge Martínez-Téllez, Director of the School of Physics, for his enthusiasm promoting the project.

\end{document}